\documentclass[natbib]{emulateapj}

\usepackage[]{natbib}

\newcommand{\mjup}{M$_{Jup}$}
\newcommand{\msun}{M$_{\odot}$}
\newcommand{\teff}{T$_{\rm eff}$}

\slugcomment{submitted to ApJL}
\shorttitle{Fomalhaut}
\shortauthors{Kenworthy et al.}
\begin{document}

\title{MMT/AO 5 micron Imaging Constraints on the Existence of Giant
Planets Orbiting Fomalhaut at $\sim$13-40 AU \altaffilmark{1}}

\author{Matthew A. Kenworthy}
\affil{Steward Observatory, The University of Arizona, 933 N. Cherry Ave., 
Tucson, AZ 85721, USA}
\email{mkenworthy@as.arizona.edu}

\author{Eric E. Mamajek}
\affil{Harvard-Smithsonian Center for Astrophysics,
Cambridge MA, 02138, USA}
\affil{Current address: University of Rochester, Department of Physics \& Astronomy, Rochester, NY,
14627-0171, USA} 

\author{Philip M. Hinz, Michael R. Meyer}
\affil{Steward Observatory, The University of Arizona, 933 N. Cherry Ave.,
Tucson, AZ 85721, USA}

\author{Aren N. Heinze}
\affil{Swarthmore College, 500 College Avenue, Swarthmore, PA 19081, USA}

\author{Douglas L. Miller, Suresh Sivanandam, Melanie Freed}
\affil{Steward Observatory, The University of Arizona, 933 N. Cherry Ave.,
Tucson, AZ 85721, USA}

\altaffiltext{1}{Observations reported here were obtained at the MMT
Observatory, a joint facility of the University of Arizona and the
Smithsonian Institution.}

\begin{abstract} A candidate $\lesssim$3 \mjup\, extrasolar planet was
recently imaged by Kalas et al. (2008) using HST/ACS at
12''.7 (96 AU) separation from the nearby ($d$ = 7.7 pc)
young ($\sim$200 Myr) A2V star Fomalhaut. Here we report results from
M-band (4.8\,\micron) imaging of Fomalhaut on 5 Dec 2006 using the
Clio IR imager on the 6.5-m MMT with the adaptive secondary mirror. Our
images are sensitive to giant planets at orbital radii comparable to
the outer solar system ($\sim$5-40 AU).  Comparing our
5$\sigma$ M-band photometric limits to theoretical evolutionary tracks
for substellar objects, our results rule out the
existence of planets with masses $>$2 \mjup\,  from $\sim$13-40 AU and objects
$>$13\,\mjup\, from $\sim$8-40 AU.
\end{abstract}

\keywords{}

\section{Introduction}

Approximately 300 extrasolar planets have been discovered,
predominantly through Doppler spectroscopy, transits, and microlensing
\citep[e.g.][]{Marcy05,Charbonneau07,Gaudi07}.  Radial velocity surveys indicate that the
frequency of gas giant planets greater
than 0.1 \mjup\, is about 15\% when
extrapolated out to 20 AU \citep{Cumming08}.
Direct imaging surveys to date place upper
limits on the frequency of gas giant planets
more massive than Jupiter at orbital radii
beyond 20 AU \citep[]{Kasper07,Lafreniere07,Nielsen08}.

 Substellar objects with
inferred masses below the deuterium-burning limit ($\sim$13\,\mjup)
have been imaged as members of young clusters
\citep[e.g.][]{Zapatero00, Luhman05}, at wide separation 
($>$tens AU) orbiting other young brown dwarfs
\citep[e.g.][]{Chauvin04, Luhman06}, and $\sim$1 \msun\, star
\citep{Lafreniere08}. Although the deuterium-burning limit
has acted as a de facto boundary between ``planets'' and ``brown
dwarfs'', these objects do not have birth certificates, and their
means of conception is a matter of conjecture. Given our knowledge
regarding parameters of protoplanetary disks, it is not
clear that any of the imaged companions with masses of $<$13 \mjup\,
could have formed {\it in situ}. Exotic scenarios for formation
at smaller radii and subsequent planet-planet scattering have been
proposed \citep[e.g.][]{Ford08,Mamajek07}.

Recently, two studies \citep{Kalas08, Marois08} announced the discovery of
what appear to represent the first unequivocal cases of exoplanets
being directly imaged and resolved around nearby stars. \citet{Marois08}
imaged three large gas giants apparently in
orbit around the young A-type debris disk star HR 8799. 
\citet{Kalas08} detect a companion $\lesssim$3
\mjup\, situated 12''.7 (96 AU) away from the bright 
A3V star Fomalhaut ($\alpha$ PsA; $V$ = 1.2 mag). Fomalhaut is a
target in our MMT/AO survey for substellar companions around nearby
intermediate mass stars and we report results of our recent observations
here.

Fomalhaut is a well-studied, nearby \citep[7.7 pc;][]{ESA97}, young
\citep[$\sim$200\, $\pm$\, 100 Myr;][]{Barrado98} main sequence
$\sim$1.95 \,\msun\, star with a debris disk system. The debris disk
system is remarkable for having been resolved in the sub-mm with JCMT
\citep{Holland03}, far-IR with {\it Spitzer}
\citep{Stapelfeldt04}, and optical with HST \citep{Kalas05}. 
The HST/ACS coronagraph images suggest that the cold dust belt is
$\sim$25 AU wide, with a sharp inner edge $\sim$133 AU from
Fomalhaut. From the eccentricity and sharpness of the inner edge of
the debris disk, \citet{Quillen06} predicted the existence of
a $\sim$0.05-0.3 \mjup\, planet with semi-major axis $a$ = 119 AU and
eccentricity of $\sim$0.1. The planet imaged by
\citet{Kalas08} has stellocentric separation of 119 AU and inferred
semi-major axis of $\approx$115 AU, in remarkable agreement with
Quillen's prediction. The mass predicted by Quillen
is lower than the upper limits derived by
\citet{Kalas08} ($\sim$1.7-3.5 \mjup), however they emphasize that
their 0.6\,$\mu$m flux
may be contaminated by an extensive circumplanetary disk.

Young giant planets are predicted to be hot (for M$_{Jup}>3$, \teff\,$>$\,300\,K for
ages $<$500 Myr \citep[]{Baraffe03}), and theoretical spectral energy
distributions (SEDs) predict a strong peak around
$\sim$5\,$\mu$m \citep[e.g.][]{Burrows97}. While direct imaging
surveys for substellar companions to nearby stars have concentrated on
near-IR bands (e.g.  $H$ and $K$), the models predict that
$L$ and $M$-band fluxes for planets should be much brighter than at
$J$, $H$, and $K$.  For example, a 10\,M$_{Jup}$ object with age
0.5\,Gyr has predicted colors of $J-M$ $\simeq$ 4, $H-M$ $\simeq$ 4,
$L-M$ $\simeq$ 1 \citep{Baraffe03}.  Motivated by these 
predictions, we initiated surveys of nearby ($d$ $<$
25 pc) stars of various types to search for substellar companions at
wide separations ($>$10 AU) using the Clio IR camera on the 6.5-m
MMT telescope with the adaptive secondary mirror \citep{Lloyd-Hart00,
Wildi03, Brusa04}. Here we report observations with MMT/AO and the
Clio IR camera sensitive to giant planets at a wide range of orbital
radii interior to the companion reported by
\citet{Kalas08}.

\section{Observations}

Fomalhaut was imaged 5 December 2006 (02:06 UT) using the Clio
3-5\,$\mu$m imager in conjunction with the adaptive secondary mirror on
the 6.5-meter MMT telescope.  The Clio detector is a high well depth
Indigo InSb detector with 320x256 pixels and 30\,$\mu$m size pixels
\citep{Hinz06}. Images were taken with a Barr
Associates M-band filter with half power wavelength range of 4.47 to
5.06 $\mu$m with central peak wavelength of 4.77 $\mu$m
\footnote{Current information on the Clio camera and MMT adaptive
secondary are available at the wiki website
http://mmtao.org/wiki/doku.php?id=mmtao:clio}.  The field of view at
$M$-band is 15''.6 $\times$ 12''.4 on the Clio array. The star was
nodded 5''.5 along the long axis of the detector after 5 images were
taken. Each of the 375 images consists of 50 coadded exposures of 209.1 ms
length, for a total coadded duration of 3920 sec. 
Exposure times were calculated so as to keep the
sky flux counts just below the nonlinearity limit for the detector
(around 40,000 ADU). Short exposures, typically of 64.1 msec were taken
after the sequence of deep exposures so as to provide photometric
check.  To avoid variations in the pattern of illumination on the
Clio detector, the instrument is fixed in orientation with respect to
the telescope, resulting in total field rotation of $21^o$ for our Fomalhaut data.  Conditions were photometric, and the
native seeing throughout the Fomalhaut imaging was $\sim$0''.5-0''.7 at an airmass of 2.1 to
2.8 (as seen with the optical acquisiton camera).


The pixel scale was determined from images of the A-type binary Castor
on UT 7 Dec 2006 \citep[orbit from ][]{Worley96}. The observed
separation at M-band was 90.93\,$\pm$\,0.06 pixels, and the orbit
predicted an angular separation of 4''.445, leading to a pixel scale
of 48.88\,$\pm$\,0.03 mas\,pixel$^{-1}$.

\begin{figure}
\epsscale{.80}
\plotone{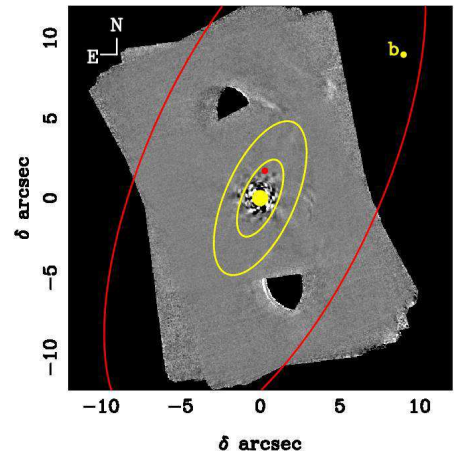}
\caption{Clio 5-micron image of Fomalhaut and environs. The location of
Fomalhaut b \citep{Kalas08} is marked at the upper right, just off the field of view of
the observations. The yellow dot marks the location of Fomalhaut. The red ellipse marks
the location of the dust belt at 140 AU \citep{Kalas05}, with the red dot marking the center of
the dust belt ellipse. The yellow ellipses represent distances
of 20 AU and 40 AU from the central star. The image is scaled linearly. No significant point
sources are detected in the image.
\label{overview}}
\end{figure}

\section{Analysis}

The Clio images were reduced using a suite of custom C routines
\citep{Heinze07_PhD} that match temporally adjacent (or
nearly adjacent) beam pair observations for background subtraction,
then rotate and coadd the background-subtracted images into final
images.  Postage stamp images of all of the individual observations
(each are typically $\sim$6-25 sec integrations) were inspected, and a
small number of observations were rejected (usually
during periods of poor seeing, or if the AO system loop was lost, or
both).  Several methods of image combination are run as separate trials
to determine the one with the most robust sensitivity, and are discussed
in Heinze (2008). A separate pipeline, written in IRAF scripts
and Perl Data Language\footnote{http://pdl.perl.org/}, provides an additional check to the other
pipelines.

We use the method of Angular Differential Imaging (ADI;
\citet[]{Marois06}) to calibrate out the presence of residual speckles in the
instrument whilst preserving the flux of any faint companions. The data
sets were split according to their beam switch and a master PSF for each
beam position created. Any faint companions at constant position angle
will be washed out in the median-combining to form the master PSF. This
PSF is then subtracted off all the individual frames. The frames are
then rotated so that North is up and East to the left, and then combined
using various sigma clipping rejection algorithms to produce the final
sensitivity image.  At small separations $(<2''.0$), the images are contrast
limited and not sky background limited, a result of the time varying
nature of the aberrations in the telescope optics.

\begin{figure}
\epsscale{.95}
\plotone{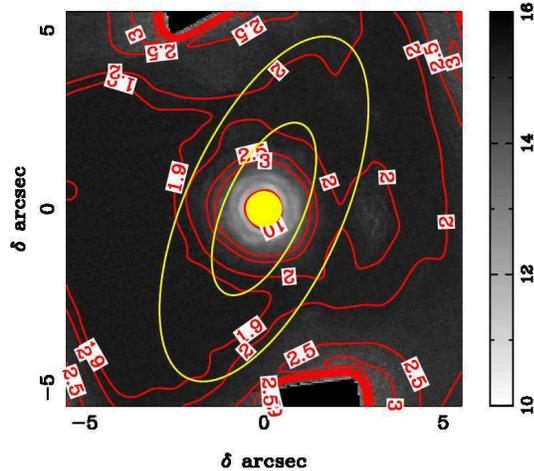}
\caption{Enlarged view of Figure \ref{overview} showing 5-sigma sensitivity
contours in Jupiter masses. For small areas we reach down to a 5-sigma
detection limit of 1.85$M_{jup}$. The underlying image shows the 5-sigma
sensitivity image in M band magnitudes marked on the scale bar.
\label{fig2}}
\end{figure}

Although Fomalhaut is an infrared photometric standard
\citep{vanderBliek96}, the core of the star's PSF in our shortest
exposure (64.1 msec) appears to be saturated. For this reason, the
photometric calibration was tied to three other A-type
stars observed that night: $\beta$ UMa, $\iota$ UMa, and $\zeta$ Lep.  The
stars do not have published ground-based M-band photometry, but they
do have published ground-based fluxes in the 1-8\,$\mu$m range
\citep{Gezari99} as well as predicted fluxes for the MSX mission 
at neighboring wavelengths \citep{Egan96}. Based on the
\citet{Gezari99} and \citet{Egan96} fluxes, we interpolate the
following M-band magnitudes for our calibrator stars: m[4.8] = 2.34
mag ($\beta$ UMa), 2.63 ($\iota$ UMa), and 3.27 ($\zeta$ Lep).
Conservative photometric uncertainties are $\pm$0.03 mag dispersion.

These three stars were observed at airmasses from 1.05 to 1.45, allowing
extrapolation of photometry to the airmass range (2.1-2.5) of the
Fomalhaut observations. A second check of the photometry was performed
by estimating Fomalhaut's flux using the unsaturated first Airy ring in
the AO corrected images.  These flux estimates agree with the airmass
extrapolation to 10\%, confirming the stability of the observing
conditions during the whole night.

Our final coadded intensity image is shown in Figure \ref{overview}. The
residuals from the ADI PSF subtraction of the star are present as a speckle
pattern in the middle of the image. The red ellipse and red dot mark the
location and center of the dust belt as imaged by \citet{Kalas05}.
The location of the exoplanet Fomalhaut b is at the upper
right of the image, just out of the field of view of the combined Clio
observations.  The combination of beamswitched images requires the
masking of the negative beam, which combined with the field rotation
leads to the two triangular areas of no coverage within the larger
image.

\begin{figure}
\includegraphics[angle=270,scale=.40]{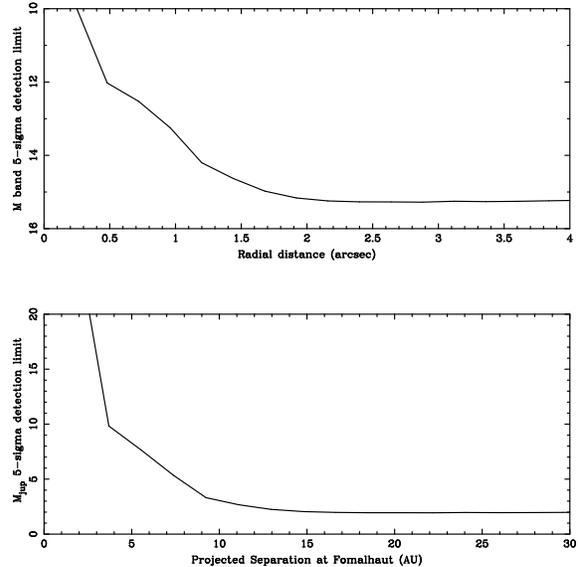}
\caption{Azimuthally averaged contrast curves for the observations of Fomalhaut. The upper
panel shows the 5-sigma point source M band magnitude detection limit as
calculated from the 375 individual Clio images. The equivalent detection
limits for COND model Jupiter mass objects is shown in the lower panel.
\label{fig3}}
\end{figure}

 For our field rotation of $21^o$, the ADI method has reduced companion
sensitivity for distances smaller than 0.86 arcseconds \citep{Marois06}.
Figure \ref{fig2} is an
expanded view of the first figure, and shows the $5\sigma$ point source
sensitivity limits assuming M band fluxes from the COND models
\citep{Baraffe03}, appropriate for temperatures $<1800K$. Sensitivity curves for regions close to Fomalhaut are
shown as azimuthally averaged plots in Figure \ref{fig3}. Although blind sensitivity tests have shown 50\% completeness at
$5\sigma$ observations \citep{Heinze07_PhD}, we keep our point source
sensitivities to be consistent with the limits quoted by other planet
surveys.

\section{Discussion}

It is unclear whether there is any connection between the presence of gas giant planets
and debris disks \citep{Moro-Martin07}.
With the discovery of a gas giant planet near the inner edge of its
outer debris belt, the question arises: does Fomalhaut have giant
planets at smaller radii?  Current analysis of the debris system
suggests inner (R $<$ 20 AU) and outer (R $>$ 100 AU) planetesimal
belts responsible for the bulk of the mid-IR and far-IR/sub-mm
emission respectively \citep{Greaves98, Stapelfeldt04, Kalas05}.
However, there is plenty of room for additional gas/ice giant planets
in the system. Our upper limits rule out masses greater than 2
\mjup\, between $\sim$13-40 AU. \citet{Chatterjee08} suggest that gas
giants could end up at large separations due to planet-planet
scattering.  In this scenario, the largest planets in the system tend
to stay put and smaller planets end up in large, eccentric orbits. Our
results suggest this may not be an explanation for the location of
Fomalhaut b.

\citet{BarnesGreenberg07} have explored the hypothesis that most
planetary systems are ``packed'' in the sense that any orbit dynamically
stable on timescales of order the age of the system or longer are
inhabited.  This implies that the planet formation process is very
efficient indeed and is consistent with numerical integration of the
orbits in our own solar system \citep{Laskar96}.  This hypothesis
found recent confirmation in the discovery of HD 74156 d
\citep{Bean08} of the mass and orbit predicted. 
If Fomalhaut has a multiplanet system spaced like the Solar system and HR
8799 systems (giant planets spaced logarithmically, spaced by
$\sim$0.25\,$\pm$\,0.05 dex in $a$), one would naively predict
interior planets at radii $\sim$65, $\sim$35, $\sim$20 AU. {\it
Spitzer} detected evidence of a warm inner disk at $<$20 AU
\citep{Stapelfeldt04}. It is tempting to speculate that Fomalhaut's
inner disk is likewise being perturbed by another planet.
If Fomalhaut indeed has a packed system of planets between its
inner ($<$20 AU) and outer ($>$133 AU) debris belts, our M-band
results suggest that planets in the $\sim$13-40 AU range
are less than $<$2 \mjup\, in mass. \citet{Chiang08} suggest that Fomalhaut exhibits an
``anomalous'' acceleration in the {\it Hipparcos} astrometry
\citep{ESA97, vanLeeuwen07}, consistent with a $\sim$30 \mjup\, brown
dwarf at $r$ $\sim$ 5 AU. Our observations rule out the existence of an
object at this mass at this orbital radius, and the data can rule out
the existence of any brown dwarfs ($>$13 \mjup) at separations of
$\sim$8-40 AU.  The accelerations in the original Hipparcos database are of low 
significance in both RA and Dec ($<$2$\sigma$), and so 
may simply be statistical noise. Additional 
observations are planned to detect the thermal emission from Fomalhaut b 
in order to further explore its properties.

\acknowledgements

We thank the CfA TAC for allocating the MMT time that made these
observations possible.  We thank the MMT staff, especially John McAfee,
Alejandra Milone, Mike Alegria, and Tim Pickering. We also thank Vidhya
Vaitheeswaran and Thomas Stalcup for their support of the MMT/AO system.
Clio is supported by grant NNG 04-GN39G from the NASA Terrestrial Planet
Finder Foundation Science Program. MAK is supported by grant NNG
06-GE26G from the NASA Terrestrial Planet Finder Foundation Science
Program. EEM was supported by a Clay Postdoctoral Fellowship during this
observing program. MRM acknowledges support through LAPLACE from the
NASA Astrobiology Institute.

\bibliography{mamajek}{}
\bibliographystyle{apj}

\end{document}